\documentclass{article}%
\usepackage{amsmath}
\usepackage{amssymb}%
\usepackage{amsfonts}%
\usepackage{graphicx}
\begin{document}

\begin{center}
{\Large Some remarks about the baffling Higgs physics}

{\Large and the particle mass problem}{\large \medskip}
\end{center}

P.\ Brovetto$^{\dag}\footnote{) Email: pbrovetto@gmail.com}$, V. Maxia$%
^\dag$\ and M. Salis$^\ddag$

$^{\dag}$On leave from Istituto Fisica Superiore - University of Cagliari, Italy

$^{\ddag}$Dipartimento di Fisica - University of Cagliari, Italy\medskip

\textbf{Abstract - }A statistical model is proposed which ascribes the Z$^{0}
$ mass to the screening properties of the neutrino Fermi sea (neutrino
vacuum). Concerning the fermion masses, some puzzling features of the Higgs
mechanism are examined. Arguments are advanced, based on the Zitterbewegung
theory of the electron substructure, showing that in low energy experiments
electron behaves as an extended distribution of charge, though its size comes
out less than 10$^{-16}$ cm in high energy experiments. This might be a clue
for explaining the origin of fermion masses without resorting to Higgs field.

\medskip

\textbf{1} \textbf{-} \textbf{Introduction}. - This year, the reconditioned
LHC at CERN in Geneva will start to operate so that experiments either
confirming or refuting the Higgs boson existence will be at last carried out.
This renewes the interest for this elusive particle which was hypothized to
explain masses of bosons which mediate the weak force and which should give
mass to all massive fermions, leptons and quarks. For this reasons, it seems
timely reconsider the argument of masses even to highlight possible
explanations different from the Higgs hypothesis.\medskip

\textbf{2} \textbf{-} \textbf{The weak boson masses}. - The main feature of
weak interaction is its very short range. In the 1933 Fermi's theory, it was
regarded as a "contact" interaction acting at zero spatial separation in
contrast with electric force mediated by photons which acts at large
distances. Higgs mechanism ascribes to weak bosons a "true" inertial mass
originated by interaction with a doublet of scalar fields in $SU(2)$ space,
that is, the Higgs field $\left[  1\right]  $. Owing to energy-time
uncertainty principle, weak bosons of mass $M$ last a time $\delta
t\leq\hslash/Mc^{2}$ so that range of weak force is $\hslash/Mc$, the boson
Compton wavelength. This is like what occurs with Yukava force mediated by
massive pions. An alternative, more conservative, explanation is based on the
effect of the neutrino Fermi sea (neutrino vacuum) on the weak boson
propagation. Indeed, owing to the vanishingly small neutrino mass, neutrino
sea is not above-bordered by a forbidden energy gap as the electron Fermi sea.
Consequently, it screens the weak force quite as electrons in the conduction
band of metals screen the electric force $\left[  2\right]  $. In this way,
range of weak force is curtailed, which is equivalent to have massive bosons.

To work out a rigorous treatment of the above screening effect is a rather
exacting task so that it appears suitable use a simplified approach. It is
based on a special application of the Thomas-Fermi method (TF), already
examined ten years ago $\left[  3\right]  $. A short account of this is given here.

\textbf{3} \textbf{-} \textbf{Screening of weak force in\ neutrino vacuum}. -
According to $SU\left(  2\right)  \otimes U\left(  1\right)  $ symmetry, weak
interactions are mediated by W$^{\pm}$ and Z$^{0}$ bosons. If mass is assigned
to these particles, their masses turn out to be related by%

\begin{equation}
M_{\text{W}^{\pm}}=M_{\text{Z}^{0}}\cos\vartheta_{w}\text{,}\label{rr}%
\end{equation}
$\vartheta_{w}\simeq$ $28.6%
%TCIMACRO{\U{b0}}%
%BeginExpansion
{{}^\circ}%
%EndExpansion
$ standing for the electroweak angle $\left[  1\right]  $. Equation (\ref{rr})
holds independently of the mass-generating mechanism. Therefore, it is
sufficient consider only the neutral Z$^{0}$ boson. In the unperturbed sea,
$\nu_{\text{L}}$ neutrinos of kinetic energies $w=cp$ ranging from $0$ to
$-\ \infty$ are present. Taking into account only one spin component, neutrino
density of states is related to kinetic energy by $\left[  4\right]  $%

\begin{equation}
\rho\left(  cp\right)  =\frac{\left(  cp\right)  ^{2}}{\pi^{2}\left(  \hslash
c\right)  ^{3}}\text{.}\label{cc}%
\end{equation}
In presence at the point $r=0$ of a steady fermion (\footnote{) We consider
here a point-like fermion since we are concerned with perturbation of Fermi
sea in space around the fermion centre. In Section \ 5 \ this issue is
re-examined.}, only the time-component $Z_{0}$ of its potential is different
from zero. So, the perturbed neutrino energy is%

\begin{equation}
w=cp+U\left(  r\right)  \text{,}\label{aa}%
\end{equation}
$U\left(  r\right)  =-\ Q_{\nu_{\text{L}}}eZ_{0}\left(  r\right)  $ standing
for the neutrino potential energy and $Q_{\nu_{\text{L}}}=1/\sin2\vartheta
_{w}$ for the electroweak neutrino charge in units of $e$. Let us examine
first the case in which $U$ is negative, that is, $U=-\left\vert U\right\vert
$. On line $w=U\left(  r\right)  $, it follows from equation (\ref{aa})
$cp=0$\ . Consequently, the neutrino sea is divided in two regions: one above
line $U\left(  r\right)  $ where $cp$ is positive, the other below this line
where $cp$ is negative. By denying that neutrino sea is perturbed up to
infinite depth (\footnote{) Perturbation in neutrino sea is defined as:
$P\left(  cp\right)  =\left\vert U\right\vert /\left(  \left\vert
cp\right\vert +\left\vert U\right\vert \right)  $. \ On line $U\left(
r\right)  $, we have $P\left(  0\right)  =1$, on the sea border $P\left(
\left\vert U\right\vert \right)  =1/2$, but \ $P\left(  cp\right)
\rightarrow0$ \ when $cp\rightarrow-\infty$.}, we assume that an energy
$w_{\text{F}}$ ($w_{\text{F}}=\left\vert w_{\text{F}}\right\vert $) can be
found great enough with respect to $\left\vert U\left(  r\right)  \right\vert
$ ($w_{\text{F}}\gg\left\vert U\left(  r\right)  \right\vert $ for any $r$)
that neutrinos with energy below $-\left[  w_{\text{F}}-\left\vert U\left(
r\right)  \right\vert \right]  $ remain unperturbed. This amounts to say that
energy $-w_{\text{F}}$\ sets a cut-off in neutrino sea depth. So, neutrino
density in the unperturbed Fermi sea is%

\begin{equation}
n_{F}=\int_{-w_{\text{F}}}^{0}\rho\left(  cp\right)  \text{d}\left(
cp\right)  =\frac{w_{\text{F}}^{3}}{3\pi^{2}\left(  \hslash c\right)  ^{3}%
}\text{.}\label{bb}%
\end{equation}
In the perturbed sea, integration over $cp$ covers the range $-\left(
w_{\text{F}}-\left\vert U\right\vert \right)  $ to $0$ in the negative region
and from $0$ to $\left\vert U\right\vert $ in the positive one. Accordingly,
the perturbed neutrino density is%

\[
n=\int_{-\left(  w_{\text{F}}-\left\vert U\right\vert \right)  }^{0}%
\rho\left(  cp\right)  \text{d}\left(  cp\right)  +\int_{0}^{\left\vert
U\right\vert }\rho\left(  cp\right)  \text{d}\left(  cp\right)  =
\]

\begin{equation}
=\frac{1}{3\pi^{2}\left(  \hslash c\right)  ^{3}}\left(  w_{\text{F}}%
^{3}-3\left\vert U\right\vert w_{\text{F}}^{2}+3\left\vert U\right\vert
^{2}w_{\text{F}}\right)  \text{.}\label{dd}%
\end{equation}
By subtracting the unperturbed density $n_{\text{F}}$, we obtain%

\begin{equation}
n-n_{\text{F}}=-\frac{\left\vert U\right\vert w_{\text{F}}^{2}}{\pi^{2}\left(
\hslash c\right)  ^{3}}\left(  1-\frac{\left\vert U\right\vert }{w_{\text{F}}%
}\right)  \text{.}\label{ee}%
\end{equation}
When potential energy $U$ is positive, that is, $U=\left\vert U\right\vert $,
equation (\ref{ee}) is found again but with term $\left\vert U\right\vert
/w_{\text{F}}$ reversed in sign $\left[  3\right]  $. In reality, taking into
account that $w_{\text{F}}$ is large with respect to $\left\vert U\left(
r\right)  \right\vert $, term $\left\vert U\right\vert /w_{\text{F}}$ is small
and can be disregarded.

By letting $Q_{f}e$ be the fermion electroweak charge, potential $Z_{0}$ turns
out to be ruled by the Poisson-like equation%

\begin{equation}
\Delta_{2}Z_{0}=-4\pi Q_{\nu_{\text{L}}}e\left(  n-n_{F}\right)  -4\pi
Q_{f}e\delta\left(  r\right)  \text{,}\label{ff}%
\end{equation}
that is, utilizing equation (\ref{ee}),%

\begin{equation}
\Delta_{2}Z_{0}-\frac{1}{\lambda_{S}^{2}}Z_{0}=-4\pi Q_{f}e\delta\left(
r\right)  \text{,}\label{gg}%
\end{equation}
where%
\begin{equation}
\lambda_{S}=\sqrt{\frac{\pi}{2}}\ \frac{\left(  \hslash c\right)  ^{3/2}%
}{Q_{\nu_{\text{L}}}ew_{\text{F}}}\text{.}\label{hh}%
\end{equation}
It follows that the screened potential is%

\begin{equation}
Z_{0}=Q_{f}e\frac{\exp\left(  -r/\lambda_{S}\right)  }{r}\text{.}\label{ii}%
\end{equation}
This result entitles us to define a "screening mass" $M_{\text{Z}^{0}}$ by
means of a formal Compton wavelength%

\begin{equation}
\frac{\hslash}{M_{\text{Z}^{0}}c}=\lambda_{S}\text{.}\label{pp}%
\end{equation}

We consider now the high-energy collisions. Let $w^{\left(  -\right)  }$,
$w^{\left(  +\right)  }$ be the energies and $\overrightarrow{p}^{\left(
-\right)  }$, $\overrightarrow{p}^{\left(  +\right)  }$ the momenta of the
colliding electron-positron pairs. Assuming momentum $\overrightarrow
{p}^{\left(  -\right)  }$ opposite to momentum $\overrightarrow{p}^{\left(
+\right)  }$, we have%

\begin{equation}
w_{c.m.}^{2}=\left(  2m_{e}c^{2}\right)  ^{2}+\left(  c\overrightarrow
{p}_{c.m.}\right)  ^{2}\text{,}\label{ll}%
\end{equation}
$w_{c.m.}=w^{\left(  -\right)  }+w^{\left(  +\right)  }$ and $\overrightarrow
{p}_{c.m.}=\overrightarrow{p}^{\left(  -\right)  }-\overrightarrow{p}^{\left(
+\right)  }$ standing for energy and momentum in the centre of mass.
Disregarding the rest energy $2m_{e}c^{2}$, wavelength corresponding to
$p_{c.m.}$ is%

\begin{equation}
\lambda_{c.m.}=\frac{\hslash}{p_{c.m.}}=\frac{\hslash c}{w_{c.m.}}%
\text{.}\label{mm}%
\end{equation}
A resonant collision is originated when wavelength $\lambda_{c.m.}$ becomes
equal to the width of the potential well which allows for the
electron-positron interaction, that is,%

\begin{equation}
\lambda_{c.m.}=\lambda_{S}\text{.}\label{nn}%
\end{equation}
So, taking into account equations (\ref{nn}), (\ref{pp}) and (\ref{mm}), we get%

\begin{equation}
w_{c.m.}=M_{\text{Z}^{0}}c^{2}\text{,}\label{oo}%
\end{equation}
which relates collision energy to Z$^{0}$ mass. When resonance occurs, energy
$w_{c.m.}$ is released through lepton and quark emissions mediated by
flavor-diagonal interactions $\left[  1\right]  $. Utilizing
equations\ (\ref{pp}) and (\ref{hh}), Z$^{0}$ mass, in energy units, turns out
to be%

\begin{equation}
M_{\text{Z}^{0}}c^{2}=\sqrt{\frac{2\alpha}{\pi}}Q_{\nu_{\text{L}}}%
w_{F}=8.11\cdot10^{-2}w_{F}\text{,}\label{qq}%
\end{equation}
$\alpha$ standing for the fine structure constant. Apart from the neutrino
electroweak charge, it is related only to $w_{F}$, the energy cut-off in
neutrino sea depth. Without this cut-off, that is, for $w_{F}\rightarrow
\infty$, the Z$^{0}$ mass diverges so reducing to zero the range $\lambda_{S}%
$\ of weak forces and recovering the old Fermi's "contact" theory.

By comparing $M_{\text{Z}^{0}}c^{2}$ with its experimental value of about $91
$ GeV, we obtain $w_{F}$ as large as $1120\ $GeV, while the screening length
$\lambda_{S}$\ turns out to be $2.2\cdot10^{-16}$cm. As for the meaning of
these figures, it is to be pointed out that on a temperature scale $w_{F}$
corresponds to $1.3\cdot10^{16}$ K. Consequenly, for $T>10^{16}$ K negative
kinetic energy neutrinos are excited to positive energies and the neutrino sea
becomes partially empty. This hinders the sea screening properties. Higgs
mechanism also gives $10^{16}$ K as the temperature which restores the
simmetry broken at low temperature $\left[  5\right]  $. This fact is not
surprising because Higgs and screening mechanisms are calibrated on equivalent
experimental data. The range $2.2\cdot10^{-16}$cm of the weak force entails
that the electron size is at least three orders of magnitude smaller than the
classic electron radius $2.8\cdot10^{-13}$cm.

Conclusion: the above statistical treatment, though not rigorous, readily
explains how massless weak bosons may originate short range interactions
without resorting to Higgs physics.\medskip

\textbf{4} \textbf{-} \textbf{The fermion masses. - }The Higgs field, which
has been assumed to be at the origin of weak boson masses, is also considered
in connection with the fermion masses. The theory accommodates the masses of
electrons and quarks of the three flavors and sets to zero neutrino masses as
a consequence of the non-existence of right neutrinos. Masses are assumed
proportional to the Higgs vacuum-expectation-value. Accordingly, three
arbitrary coupling factors $g_{e,}$ $g_{u}$, $g_{d}$ are considered for the
first flavor. So, taking into accont the second and third flavor, the theory
contains nine undetermined parameters $\left[  1\right]  $.

To detect Higgs boson, various kinds of experiments have been devised based on
its decay. But, the mere existence of a decay showing the features expected
for the Higgs is not sufficient to conclude that it really concerns the "true"
Higgs. It is also necessary that the found results allow an independent
determination of the above mentioned nine parameters. Obviously, in lack of
this, the Higgs remains nothing more than a conjecture.

But, another point challenges the correctness of the Higgs hypothesis. It is a
well-established result of classic electrodynamics that the electromagnetic
(e.m.) field shows inertial properties. This field, indeed, shows a momentum
density $\left(  \overrightarrow{E}\times\overrightarrow{H}\right)  /4\pi c$
parallel to the Pointing's propagation vector. On this ground, at the
beginning of the past century considerable endeavour was devoted to explain
the electron mass as the e.m. mass of a charge distribution of definite size.
The advent of quantum mechanics set an end to these attempts, because the
supposed point-like nature of elementary particles entails a divergent
electron mass. For this reason, the unsolved problem of electron mass was
merely put aside by applying renormalization procedures purposely devised. It
follows that a viable Higgs mechanism, besides the mass, should likewise
explain how the divergent electron e.m. mass is turned off.

In our opinion, in order to manage the tough problem of particle masses, two
basic arguments should be considered. The first is that, according to the
Copenhagen interpretation of quantum physics, electron is an observable
object, not an absolute entity. This means that its features, as expected from
theory, depend on the special experiment considered. The second concerns a
peculiar property of Dirac equation: the so-called electron Zitterbewegung
(Zb) $\left[  6,7,8\right]  $. It has been shown, when dealing with the
expected electron velocity, that in the Fourier expansion of the spinor
components, each element $dp_{x}dp_{y}dp_{z}$ of momentum space is associated
with oscillations on $x$, $y,$ $z$ axes of amplitudes and phases depending on
$\overrightarrow{p}$. These oscillations are caused by interference beats
between positive and negative energy states. By integrating over momenta, we
have for $x$ axis,%

\begin{equation}
\left\langle x\left(  t\right)  \right\rangle _{\text{Zb}}=\frac{\lambda_{C}%
}{4\pi}%
%TCIMACRO{\diiint }%
%BeginExpansion
{\displaystyle\iiint}
%EndExpansion
A_{1}\sin\left(  4\pi\frac{t}{T_{\text{Zb}}}+\varphi_{1}\right)
d^{3}\overrightarrow{p}\text{,}\label{ss}%
\end{equation}
$\lambda_{C}$ standing for the Compton wavelength and%

\begin{equation}
T_{\text{Zb}}=\frac{h}{m_{e}c^{2}}=8.1\cdot10^{-21}\text{s}\label{tt}%
\end{equation}
for twice the oscillation period. It follows that a dynamic substructure is
originated in which electron move in space around its centre of mass along a
complex tissue of closed trajectories $\left[  9\right]  $.

Keeping the above arguments in mind, we point out that high-energy collisions
are very fast processes. The collision time $\tau$ can be roughly identified
with the ratio between impact parameter $b$ and the electron-positron
velocities $c$, that is, $\tau\simeq b/c$ $\left[  10\right]  $. But, while
impact parameter in direction orthogonal to velocities can be assumed equal to
the electron size, that is, $b_{\perp}\simeq10^{-16}$cm, in parallel direction
it is reduced by Lorentz contraction, that is, $b_{\Vert}\simeq\sqrt
{1-\beta^{2}}\cdot10^{-16}$cm. Since for $45$ GeV electrons we have:
$\sqrt{1-\beta^{2}}=1.1\cdot10^{-5}$, we obtain $\tau\simeq b_{\Vert
}/c=3.7\cdot10^{-32}$s. This time is short in comparison with the Zb period,
in fact: $\tau/T_{\text{Zb}}=4.6\cdot10^{-12}$. It follows that when electrons
collide oscillations are stopped and the\ Zb substructure is not observable.
This is like what occurs with a high-speed camera which allows to take steady
pictures of a propeller even if it spins very fast.

The situation is opposite in low energy experiments where, in general,
energies are determined with high accuracies. For instance, in atomic
spectroscopy indetermination $\delta w\ $is less than about $10^{-7}%
%TCIMACRO{\unit{eV} }%
%BeginExpansion
\operatorname{eV}
%EndExpansion
$, that is: $\delta w/m_{e}c^{2}\lesssim2\cdot10^{-13}$. This follows from the
fact that Rydberg constant (R$_{\text{H}}=13.6056981\
%TCIMACRO{\unit{eV}}%
%BeginExpansion
\operatorname{eV}%
%EndExpansion
$) is known with seven decimal digits. Considering that equation (\ref{tt})
allows us to write the energy-time uncertainty principle $\delta w\ \delta
t\simeq h $ as a reciprocity relation%

\begin{equation}
\frac{\delta w}{m_{e}c^{2}}\frac{\delta t}{T_{\text{Zb}}}\simeq1\text{,}%
\label{uu}%
\end{equation}
we obtain $\delta t/T_{\text{Zb}}\gtrsim5\cdot10^{12}$. This large
indetermination in time compels us to eliminate time in equation (\ref{ss}) so
that oscillations are changed into distributions of probability lying along
the electron trajectories (\footnote{) This is like what occurs with a classic
oscillator: $x=Asin\left(  2\pi t/T\right)  $. Probability that the
oscillating particle is found betveen $x$ and $x+dx$ is: $dP=2dt/T$, \ $dt$
standing for the time required to cross $dx $. We get thus: $dP/dx=2/\left(
\overset{\cdot}{x}T\right)  $, that is, $dP/dx=1/\left(  \pi\sqrt{A^{2}-x^{2}%
}\right)  $.}. Consequently, the observable electron turns out to be a static
distribution of charge of definite size. In this way, it might allow for a
finite e.m. mass evading divergent results $\left[  11\right]  $. Opposite to
the previous propeller example, this is like what occurs with a low-speed
camera which takes the picture of the spinning propeller in form of an uniform
disk.\medskip

\textbf{5} \textbf{-} \textbf{Final remarks}. - The opinion that the Zb
electron substructure should be considered in connection with the mass problem
is not new $\left[  9\right]  $. So far, however, it has found scarce
attention because most physicists consider the Zb oscillations as a
meaningless feature of Dirac's equation and assume that electron behaves
always as a point-like object which, in absence of external forces, cannot
change its own velocity. Recently, an experiment has been performed showing
clear evidence adverse to this belief $\left[  12\right]  $. In this
experiment, a single $^{40}$Ca$^{+}$ ion trapped in an electromagnetic cage
simulates a free electron in an extremely fast quivering motion superimposed
on a slow drift, that is, just the Zb motion.

It is to be pointed out , on the other hand, that the TF\ statistical model,
just in order to allow for the finite mass of weak bosons, rules out the
assumption of a point-like electron devoid of a dynamic substructure. In fact,
this electrons would originate, a divergent neutrino potential energy:
$\left\vert U\left(  r\right)  \right\vert \rightarrow\infty$ for
$r\rightarrow0$, which according to equation (\ref{aa}) would prevent us from
considering a finite cut-off energy $w_{F}$ and, consequently, a finite boson
mass. \medskip

\textbf{References}

$\left[  1\right]  $ G. Kane, Modern Elementary Particle Physics
(Addison-Wesley, 1987).

$\left[  2\right]  $ N. H. March, Philos. Mag. Suppl. \textbf{6} (1957) and
references therein.

$\left[  3\right]  $ P. Brovetto, V. Maxia and M. Salis, Il Nuovo Cimento
\textbf{A 112, }531 (1999).

$\left[  4\right]  $ L.\ D.\ Landau and E.\ M.\ Lifshitz, Statistical Physics
(Pergamon Press, Oxford, 1969).

$\left[  5\right]  $ S. Weinberg, Phys. Rev. \textbf{D 9}, 3357 (1974).

$\left[  6\right]  $ P. Dirac, The Principles of Quantum Mechanics (Oxford,
1958) \S \ 69.

$\left[  7\right]  $ B. R. Holstein, Topics in Advanced Quantum Mechanics
(Addison-Wesley, 1992).

$\left[  8\right]  $ P. Brovetto, V. Maxia and M. Salis,
arXiv:quant-ph/0702112v 12 Feb 2007.

$\left[  9\right]  $ A. O. Barut and A. J. Bracken, Phys. Rev. \textbf{D 23},
2454 (1981).

$\left[  10\right]  $ E. Fermi, Nuclear Physics (University of Chicago Press,
1955) Ch. II.

$\left[  11\right]  $ P. Brovetto, V. Maxia and M. Salis,
arXiv:quant-ph/0512047v1 6 Dec 2005.

$\left[  12\right]  $ R, Gerritsma, G. Kirchmair, F. Z\"{a}ringer, E. Solano,
R. Blatt and C. F. Ross, Nature \textbf{463}, 68 (2010).

\end{document}